\documentclass[12pt]{iopart}
\usepackage{iopams}
\usepackage{graphicx}
\usepackage{psfig}

\begin{document}

\title{PHENIX Direct Photons in 200 GeV p+p and Au+Au Collisions}

\author{Justin Frantz\dag for the PHENIX Collaboration
\footnote[3]{For the full PHENIX Collaboration author list and
acknowledgments, see Appendix "Collaborations" of this volume. }}

\address{\dag Columbia University}

\begin{abstract}

We present the first positive direct photon ($\gamma$) results in
Au+Au at $\sqrt{s_{NN}}$ = 200 GeV along with initial p+p results at
the same energy. The p+p result is found to be consistent with NLO
perturbative QCD (pQCD) predictions within its large
uncertainties.  In central Au+Au collisions, an excess over
expected background as large as 200-300$\%$ is observed from
transverse momentum ($p_T$) 4-12 GeV/c.  This large signal is
shown to be consistent with the scaled pQCD $\gamma$ prediction,
together with suppression of meson background sources.

\end{abstract}




\section{Introduction}

The study of hard scattering processes is a large and important
part of the RHIC program.  So far only hadronic production has
been measured, but hard prompt photons are anticipated to provide
complementary information.

Hard $\gamma$ production is dominated by gluon Compton scattering
where an incoming quark and gluon interact to yield a photon and
another quark.  While this process has the same initial state as
other hard parton-parton interactions, its final state has an
outgoing $\gamma$ which suffers no fragmentation and thus carries
information that is cleaner and less complicated than jet-forming
color charged counterparts. Hence, hard $\gamma$'s provide an
excellent probe to study nucleon gluon structure in p+p
collisions. Furthermore in A+A collisions, the photons should not
be distorted by the dense color fields of any media formed, once
produced.

Direct photon measurements are considered difficult because of a
large background of meson decay photons relative to signal.  This
background peaks at $p_T$ below $\sim$4GeV and makes measurements
there most difficult.  Consequently, our Au+Au results are not
currently precise enough to distinguish the expected thermal
component \cite{peitzmann} above this background or even newly
proposed sources of $\gamma$ enhancement \cite{Fries:2002kt} since
each are expected to be significant only in similar low $p_T$
regions as the background. At higher $p_T$ however, the already
observed suppression of background mesons should result in a large
intrinsic signal to background, if the hard $\gamma$ are indeed
unmodified.  This would be interesting by itself, and would also
support the claim that jet quenching suppression is an effect of
the final, and not the initial state.

\section{Data analysis and results}

The data shown here was taken in the 2001 RHIC run 2 with the
standard PHENIX setup as in \cite{Adler:2003qi} and
\cite{Reygers:2002kc}. Especially important to this analysis are:
1) the Electromagnetic Calorimeter (EMCal) which detects photon
showers and 2) The use of EMCal high $p_T$ triggers.  For this
analysis, a new set of Au+Au data corresponding to 55M sampled
events, obtained with a high $p_T$ trigger that had $100\%$
efficiency for $p_T \geq$ 5 GeV/c, was combined with the 30M
minimum bias events of \cite{Adler:2003qi}, while the p+p data set
consisted of 35M triggered events. \cite{Adler:2003qi} and
\cite{Reygers:2002kc} also describe the underlying $\pi^{0}$
analyses as well other details of the statistical direct photon
search technique we employ for this analysis.

To measure the total inclusive single photon spectrum, EMCal
showers are counted in each $p_T$ bin, with shower shape and
timing cuts applied to reduce the hadronic shower contributions.
The remaining charged track contamination is estimated by
measuring distance correlations from Pad Chamber charge particle
veto hits. $n/\overline{n}$ contamination is estimated from the
PHENIX measurements of $p/\overline{p}$ (with isospin feed-down
modifications) and a full GEANT simulation using FLUKA. Total
hadron contamination to the photon spectra peaks at 18-25$\%$ for low
values of $p_T$ ($< \sim$3 GeV) due to the contracted hadronic
shower response of the calorimeters, and is estimated with a
$p_T$-independent upper limit of 5-15$\%$ above $\sim$6 GeV/c. In
p+p, we make a further cut for background reduction which removes
$\pi^{0}$/$\eta\rightarrow\gamma\gamma$ photons in $\gamma$ pairs
with invariant mass near the $\pi^{0}$ or $\eta$ mass. Corrections
for acceptance, energy smearing, detector multiplicity skew,
conversion loss, and other effects then follow.

The expected background from photonic meson decay to this
inclusive spectrum is calculated using simple kinematic decay
simulations.  To estimate the contributing meson sources, $m_T$
scaling based on $\pi^{0}$ is used with normalization ratios from
high $p_T$ world averages \cite{plb:alb95} and additionally for
$\eta$ (0.55), from our corroborating preliminary measurements.
The ratio of measured to expected background photons is made such
that an excess $>$ 1 is interpreted as the direct photon
contribution. In order to cancel certain systematic errors, we
divide by the final $\pi^{0}$ spectrum both the photon measurement
(using $\pi^{0}$ data points) and background expectation (using a
$\pi^{0}$ fit). This makes a double ratio interpreted in the same
way.

Results are shown in the figures (1,2,3).  The major systematic
uncertainties come from the efficiency determination (10$\%$), the
$\pi^{0}$ peak extraction (9$\%$), and the hadron contamination
(12-7$\%$). For p+p, the excess ratio is then applied back to the
inclusive photons creating the invariant spectrum of direct
photons. All results are compared to pQCD direct $\gamma$
predictions \cite{Catani:1999hs}, scaled by nuclear thickness
("$N_{coll}$") and combined with our decay background calculations
for Au+Au.

\section{Results and Conclusion}

A large and significant photon excess above meson decay
background, attributable to direct photons, is found in central
Au+Au events.  The very high intrinsic signal to background ratio
at the highest $p_T$ is a new phenomenon.  The size of the signal
is found to be consistent with the nuclear thickness-scaled pQCD
expectation, given the previously observed level of suppression for
mesons at RHIC \cite{Adler:2003qi}.  Since peripheral events lack
the same suppression, the signal in these events becomes
undetectable within the current uncertainties.  The invariant
spectrum of direct photons obtained for p+p collisions is also
found to be qualitatively consistent with the same expectation
within its large uncertainties.

These observations imply the presence of a medium in central Au+Au
collisions which quenches hard partons but does not affect hard
photons.  This is exactly the behavior predicted for color dense
matter like the QGP \cite{Wang:1996pe}.  Since hard production
rates for mesons and photons are determined by the initial state,
the photon observations, like those from studies of d+Au hard
probes, indicate that such a medium exists only in the final
state.  Future PHENIX work in making the $\gamma$ measurements
more precise at lower $p_T$ values will be able to address other
QGP predictions such as modified thermal production and other
enhancement.\linebreak


\begin{figure}
    \begin{center}
        \centerline{\includegraphics[height=6cm, width=18cm
        ]{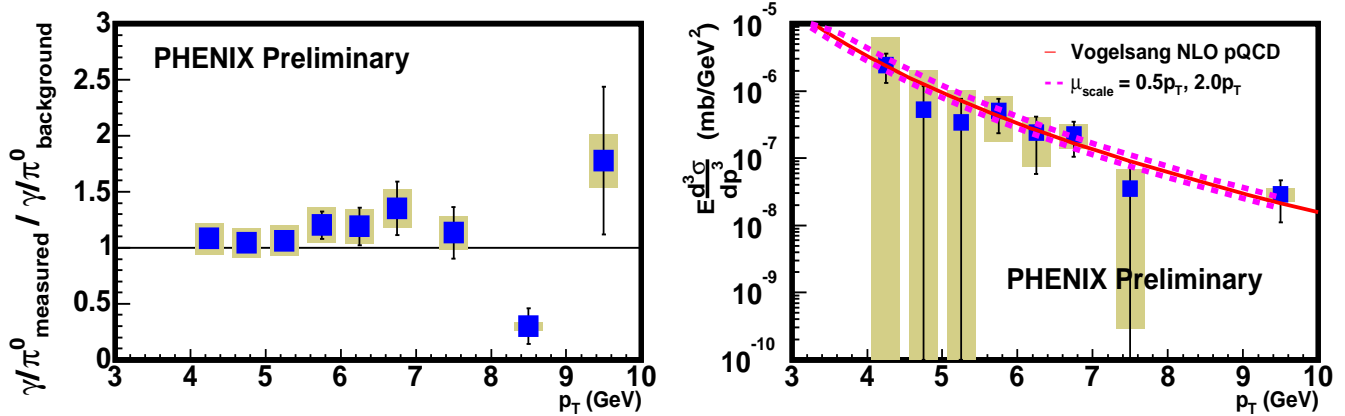}} \caption{Direct $\gamma$ excess
        double ratio and cross section in 200 GeV p+p collisions.
        The cross section is compared to pQCD
        \cite{Catani:1999hs} for scale factors $\mu$ = ${0.5p_T,
        1.0p_T, 2.0p_T}$}
         \label{fig:klaus_pp}
    \end{center}
\end{figure}

\begin{figure}
    \begin{center}
        \centerline{\includegraphics[height=9cm, width=13.5cm ]{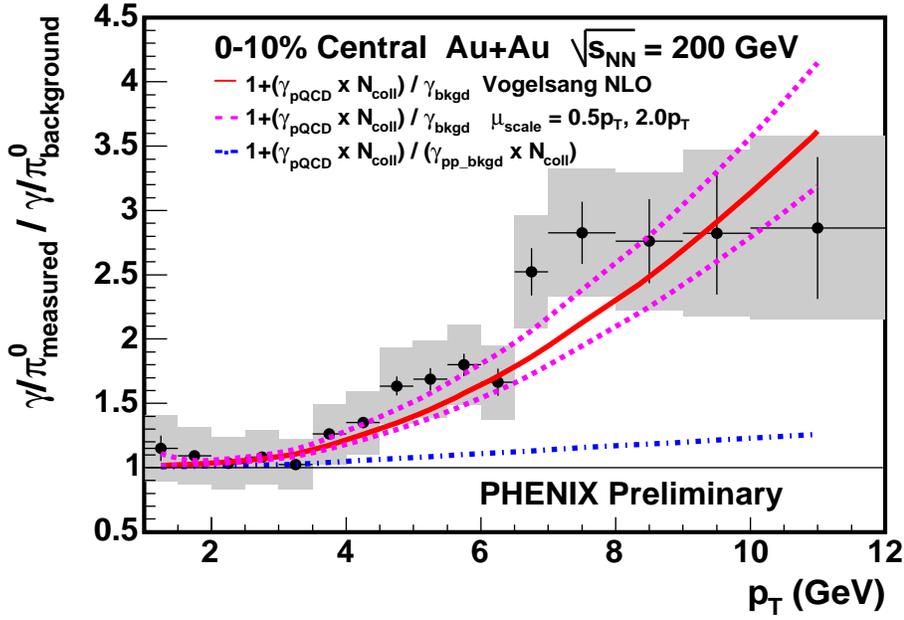}}
        \caption{Photon excess double ratio for the most central Au+Au bin, compared to
            pQCD (see text) for different scale factors as
        indicated.  The dot-dash curve represents the
        expected excess if there were no suppression of the meson
        decay background.  All uncertainties are expressed in the
        vertical error bars.  Horizontal error bars represent only
        the bin width.} \label{fig:cent0_dirgam}
    \end{center}
\end{figure}

\begin{figure}
    \begin{center}
        \centerline{\includegraphics[height=9.5cm, width=11cm ]{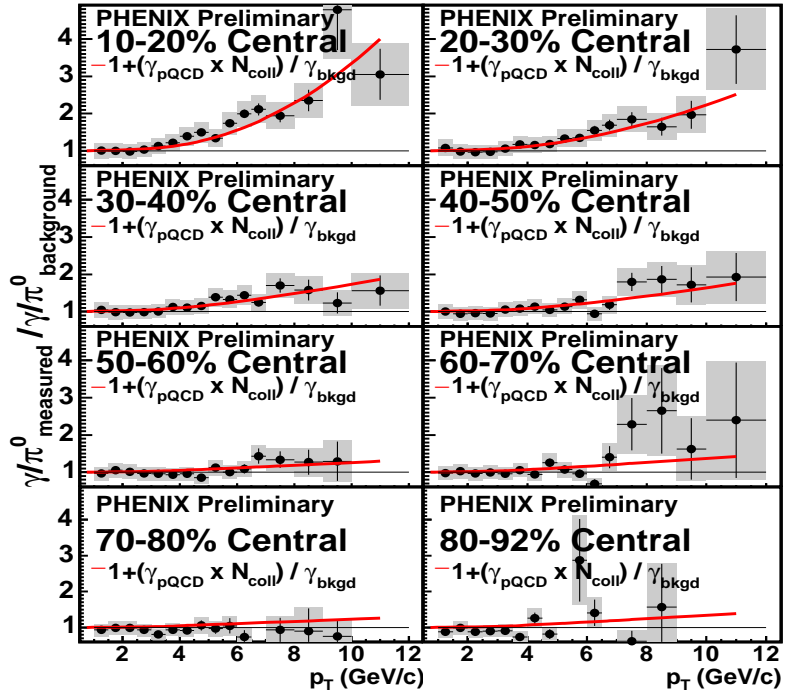}}
        \caption{Photon excess double ratio for remaining Au+Au
        centralities compared to pQCD as in Fig. \ref{fig:cent0_dirgam}}
        \label{fig:all_cent_dirgamma}
    \end{center}
\end{figure}

\end{document}